\documentclass[journal,onecolumn]{IEEEtran}

\ifCLASSINFOpdf
\else
\fi
\usepackage{url}

\usepackage[utf8]{inputenc}

\usepackage{cite}
\usepackage{hyperref}
\usepackage{support-caption}
\usepackage[subrefformat=parens,labelformat=parens]{subcaption}

\usepackage{amsmath,amssymb}
\usepackage{mathtools}
\usepackage{braket}
\usepackage{marvosym}

\usepackage{fancyhdr}

\usepackage{xcolor}

\usepackage{float}
\usepackage{stfloats}
\usepackage{siunitx}
\usepackage{booktabs}
\usepackage{multirow}
\usepackage{calc}

\usepackage{tikz}
    \usetikzlibrary{quantikz}
    \usetikzlibrary{shapes}
    \usetikzlibrary{arrows}
    \usetikzlibrary{positioning}
\usepackage{pgfplots}
    \pgfplotsset{every tick label/.append style={font=\footnotesize},compat=1.17}
\usepackage{tikzit}

\newcommand{\inv}[1]{#1^{-1}}
\newcommand{\Hs}{\mathcal{H}}
\newcommand{\Cs}{\mathbb{C}}

\begin{document}

\title{Efficient parallelization of quantum basis state shift}
%
%
%

\author{Lj. Budinski, O. Niemimäki, R. Zamora-Zamora, and V. Lahtinen \thanks{This research was partially supported by the Business Finland project 9820/31/2022 Quantum-Native Multiphysics.}
\thanks{The authors are with Quanscient Oy (Tampere, Finland) (e-mail: ljubomir.budinski@quanscient.com, ossi.niemimaki@quanscient.com, roberto.zamora-zamora@quanscient.com, valtteri.lahtinen@quanscient.com}}

\markboth{Efficient parallelization of quantum basis state shift}{}

\maketitle

\pagestyle{fancy}
\fancyfoot[L]{\copyright 2023 IOP Publishing Ltd}

\begin{abstract}
Basis state shift is central to many quantum algorithms, most notably the quantum walk. Efficient implementations are of major importance for achieving a quantum speedup for computational applications. We optimize the state shift algorithm by incorporating the shift in different directions in parallel. This provides a significant reduction in the depth of the quantum circuit in comparison to the currently known methods, giving a linear scaling in the number of gates versus working qubits in contrast to the quadratic scaling of the state-of-the-art method based on the quantum Fourier transform. For a one-dimensional array of size $2^n$ for $n > 4$, we derive the total number of $15n + 74$ two-qubit $CX$ gates in the parallel circuit, using a total of $2n-2$ qubits including an ancilla register for the decomposition of multi-controlled gates. We focus on the one-dimensional and periodic shift, but note that the method can be extended to more complex cases.
\end{abstract}


\section{Introduction}
\label{sec:introduction}
\IEEEPARstart{Q}{uantum} basis state shift is a fundamental subroutine of many quantum computing applications. In the present work we outline a new implementation for the one-dimensional state shift algorithm with two-qubit gate efficiency surpassing the known methods in the literature.

We define the state shift as an increment/decrement operation on the binary-encoded canonical computational basis states. In other words, we are interested in how the \emph{amplitudes} attached to specific quantum states can be efficiently shifted on the array defined by ordering the states based on their binary-encoded values. For clarity, we focus on the simple one-dimensional interpretation of the shift movement -- naming the possible directions \emph{left} and \emph{right} -- and consider taking steps of size one. This process can be generalized, for example by interpreting the state space as sites on a multidimensional lattice or nodes on a graph, and instead of incremental steps one could also consider larger ``jumps'' over the space. We will extend this topic in a forthcoming paper.

Our main motivation is the use of state shift as a superposition of movements as it appears in \emph{quantum random walks}. The quantum walk extends the classical random walks by considering a movement of a quantum particle entangled to a control state representing a scattering process. As a quantum computing algorithm, the quantum walk has been shown to lead to an exponential speedup for a number of graph-traversing problems; see \cite{Venegas_Andraca_2012} for a review on the topic. Aside the pure quantum walk algorithm, we outline a couple of other applications in Section~\ref{sec:applications}.

We present here a parallelization of the canonical implementation of the state shift based on multi-controlled $X$ gates as presented in \cite{Douglas_Wang}. Our algorithm introduces a \emph{parallel shift} by dividing the quantum state into \emph{even} and \emph{odd} basis components, which makes the shift more efficient in terms of gate count. The difference becomes especially pronounced in the case we can introduce ancillary qubits to decompose the multi-controlled gates: then the number of $CX$ gates can be shown to have linear scaling with respect to the number of working qubits involved, thus giving a logarithmic scaling over the number of the array states.

We compare this parallelization to the canonical version as well as to a variant based on the quantum Fourier transform proposed in \cite{Shakeel}, and provide a complexity analysis in terms of single- and two-qubit gates.

\section{Quantum state shift}
\label{sec:stateShift}

We introduce the quantum basis state shift through the lens of quantum walks, as to our knowledge it is the most fundamental algorithm that uses the state shift as its main component, and it illustrates well the possibilities for parallel computation. More applications are noted in Section~\ref{sec:applications}.

\subsection{One-dimensional state shift}

Discrete-time quantum walk is characterized by a particle moving on a grid based on a scattering process. For simplicity we assume that the grid is one-dimensional and periodically finite, making it possible to encode it with a finite state space of dimension $N \coloneqq 2^n$. At each time step of the algorithm, the particle moves either left or right, and this is mapped into either decrement or increment of the binary-encoded state in the canonical computational basis of a qubit system. Due to the periodicity, this shift is cyclical and the position is defined modulo $N$.

\begin{figure}[ht]
    \centering
    \input{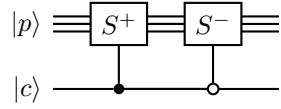}
    \caption{Increment and decrement operators subject to the control state $\ket{c}$. For simplicity the control is represented by a one-qubit state.}
    \label{fig:circuit-easy_shift}
\end{figure}

We define the basis state shift as a superposition of the increment $S^+$ and decrement $S^-$, which on their own map a binary-encoded $n$-qubit state $\ket{k}$ as $S^{\pm} \ket{k} = \ket{k\pm 1}$. A non-trivial movement is induced with a control state $\ket{c}$, leading to the state shift as a sequence of controlled increment and decrement operators: 
\begin{equation}
    \sum_k \alpha_k \ket{k}\otimes \ket{c} \mapsto S_c^-S_c^+ \left(\sum_k \alpha_k \ket{k}\otimes \ket{c} \right) .
\end{equation}
In the usual quantum walk applications the control state represents the outcome of the quantum coin flip, and it can be thought as the grid-velocity $v \coloneqq \pm 1$ of the particle with respect to the one-step movement on the line. In what follows we will call this the \emph{coin state}; note that this does not necessarily mean a uniform distribution of the probabilities, but simply indicates how the two directions are in superposition. We identify the increment with the \emph{right shift}, and the decrement with the \emph{left shift}. The overall process is outlined in Figure~\ref{fig:circuit-easy_shift}.

\begin{figure*}[ht]
    \centering
    \input{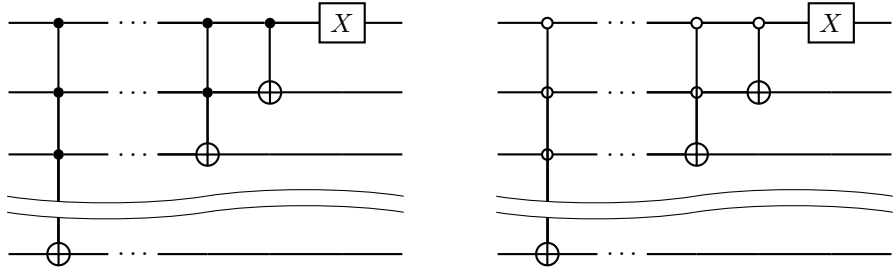}
    \caption{Shift unitaries as multi-controlled $X$ gates: increment $S^+$ followed by decrement $S^-$. We are assuming that the top qubit is the least significant in binary encoding.}
    \label{fig:circuit-shifts}
\end{figure*}

The unitaries governing the shift operators can be formed with a cascade of multi-controlled $X$ gates~\cite{Douglas_Wang}, as depicted in Figure~\ref{fig:circuit-shifts}. In Section~\ref{sec:parallel} we improve the performance of this canonical state shift by parallelizing the sequence of the increment and decrement operators.

\subsection{The QFT variation}

An interesting variation of the canonical state shift can be constructed using the quantum Fourier transform (QFT). This idea is based on the diagonalization of the increment as 
\begin{equation}
    S^+ = \inv{F}PF ,
\end{equation}
where $P$ is a diagonal phase multiplication matrix and $F$ is the $N$-dimensional QFT. For an $n$-qubit system, the operator $P$ can be given as a tensor product of $n$ single-qubit phase rotations.~\cite{Shakeel} 

This QFT variant is especially appealing on the lower qubit numbers, as the resultant circuits stay relatively simple. However, the number of two-qubit gates grows quadratically with respect to the number of working qubits; the circuit given in~\cite{Shakeel} for the complete state shift requires $n^2 + 4n + 1$ two-qubit gates when $n$ is the number of working qubits.

\section{Parallel shift}
\label{sec:parallel}

We can improve the canonical state shift algorithm by parallelizing the sequence of the increment and decrement operators. The benefit lies in removing redundant multi-controlled gates, which would otherwise hamper the scaling in terms of the grid size. In this section we explain the parallelization process, and then explore in detail the circuit complexity in Section~\ref{sec:complexity}.

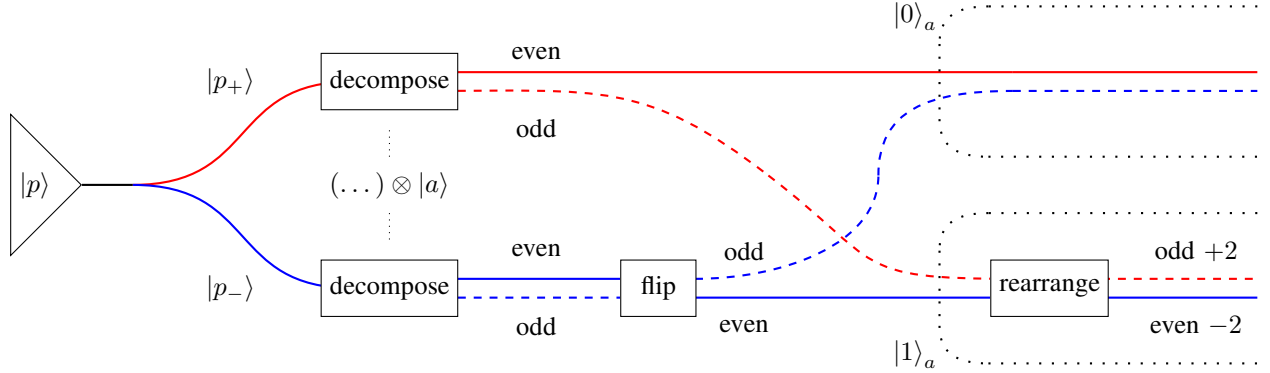
\begin{figure*}[ht]
    \centering
\tikzstyle{process box}=[fill=white, draw=black, shape=rectangle, tikzit category=boxes, minimum width=1cm, minimum height=0.75cm]
\tikzstyle{state in}=[fill=white, draw=black, shape={{}}, isosceles triangle, isosceles triangle apex angle=90, minimum size=0.5cm, shape border rotate=90]
\tikzstyle{state out}=[fill=white, draw=black, shape={{}}, isosceles triangle, isosceles triangle apex angle=90, minimum size=0.5cm, shape border rotate=270]

\tikzstyle{dash line}=[-, loosely dotted, thick, opacity=0.75]
\tikzstyle{red line}=[-, color=red, thick]
\tikzstyle{blue line}=[-, color=blue, thick]

\begin{tikzpicture}[xscale=0.65, yscale=0.5]
	\begin{pgfonlayer}{nodelayer}
		\node [style=none] (0) at (4, 19.5) {};
		\node [style=none] (3) at (10, 22) {};
		\node [style=none] (4) at (10, 17) {};
		\node [style=none] (5) at (10, 22.5) {};
		\node [style=none] (6) at (10, 16.5) {};
		\node [style=none] (7) at (22, 22.5) {};
		\node [style=none] (8) at (22, 16.5) {};
		\node [style=none] (9) at (15.5, 17) {};
		\node [style=none] (10) at (15.5, 16.5) {};
		\node [style=none] (11) at (8.5, 22.25) {};
		\node [style=none] (12) at (4, 19.5) {};
		\node [style=none] (14) at (8.5, 16.75) {};
		\node [style=process box] (15) at (9.25, 22.25) {decompose};
		\node [style=process box] (16) at (9.25, 16.75) {decompose};
		\node [style=state in, shape border rotate=0] (17) at (2, 19.5) {$\ket{p}$};
		\node [style=none] (18) at (27, 22.5) {};
		\node [style=process box] (19) at (22.75, 16.75) {rearrange};
		\node [style=none] (21) at (23.5, 17) {};
		\node [style=none] (22) at (23.5, 16.5) {};
		\node [style=none] (23) at (23.5, 16.5) {};
		\node [style=none] (24) at (27, 17) {};
		\node [style=none] (25) at (27, 16.5) {};
		\node [style=none] (26) at (22, 17) {};
		\node [style=none] (27) at (22, 22) {};
		\node [style=none] (28) at (27, 22) {};
		\node [style=none] (29) at (6, 22.25) {$\ket{p_+}$};
		\node [style=none] (30) at (6, 16.75) {$\ket{p_-}$};
		\node [style=none] (31) at (12.25, 23) {even};
		\node [style=none] (32) at (12.25, 21) {odd};
		\node [style=none] (33) at (12.25, 17.75) {even};
		\node [style=none] (34) at (12.25, 15.75) {odd};
		\node [style=none] (35) at (25.75, 17.75) {odd $+2$};
		\node [style=none] (36) at (25.75, 15.75) {even $-2$};
		\node [style=process box] (42) at (14.75, 16.75) {flip};
		\node [style=none] (43) at (14, 17) {};
		\node [style=none] (44) at (14, 16.5) {};
		\node [style=none] (47) at (16.5, 15.75) {even};
		\node [style=none] (48) at (16.5, 17.75) {odd};
		\node [style=none] (49) at (9.25, 21) {};
		\node [style=none] (50) at (9.25, 18) {};
		\node [style=none] (51) at (9.25, 19.5) {$(\dots) \otimes \ket{a}$};
		\node [style=none] (52) at (9.25, 18.75) {};
		\node [style=none] (53) at (9.25, 20.25) {};
		\node [style=none] (54) at (17, 19.75) {};
		\node [style=none] (55) at (19.25, 19.75) {};
		\node [style=none] (59) at (27, 20.25) {};
		\node [style=none] (60) at (27, 24.25) {};
		\node [style=none] (61) at (21.5, 24.25) {};
		\node [style=none] (62) at (20.5, 23.25) {};
		\node [style=none] (63) at (20.5, 21.25) {};
		\node [style=none] (64) at (21.5, 20.25) {};
		\node [style=none] (65) at (20, 24) {$\ket{0}_a$};
		\node [style=none] (66) at (27, 14.75) {};
		\node [style=none] (67) at (27, 18.75) {};
		\node [style=none] (68) at (21.5, 18.75) {};
		\node [style=none] (69) at (20.5, 17.75) {};
		\node [style=none] (70) at (20.5, 15.75) {};
		\node [style=none] (71) at (21.5, 14.75) {};
		\node [style=none] (72) at (20, 15) {$\ket{1}_a$};
	\end{pgfonlayer}
	\begin{pgfonlayer}{edgelayer}
		\draw [style=red line, in=180, out=0] (5.center) to (7.center);
		\draw [style=blue line, in=180, out=0, looseness=1.50] (10.center) to (8.center);
		\draw [style=red line, in=180, out=0, looseness=1.25] (12.center) to (11.center);
		\draw [style=blue line, in=180, out=0, looseness=1.25] (12.center) to (14.center);
		\draw [thick] (17) to (12.center);
		\draw [style=red line] (7.center) to (18.center);
		\draw [style=red line, dashed] (21.center) to (24.center);
		\draw [style=blue line] (23.center) to (25.center);
		\draw [style=blue line, dashed] (27.center) to (28.center);
		\draw [style=blue line] (4.center) to (43.center);
		\draw [style=blue line, dashed] (6.center) to (44.center);
		\draw [dotted] (53.center) to (49.center);
		\draw [dotted] (52.center) to (50.center);
		\draw [style=red line, dashed, in=135, out=0, looseness=1.25] (3.center) to (54.center);
		\draw [style=red line, dashed, in=-180, out=-45, looseness=1.50] (54.center) to (26.center);
		\draw [style=blue line, dashed, in=270, out=0] (9.center) to (55.center);
		\draw [style=blue line, dashed, in=180, out=90, looseness=1.25] (55.center) to (27.center);
		\draw [style=dash line] (60.center) to (61.center);
		\draw [style=dash line, bend right=45, looseness=1.25] (61.center) to (62.center);
		\draw [style=dash line] (62.center) to (63.center);
		\draw [style=dash line, bend right=45, looseness=1.25] (63.center) to (64.center);
		\draw [style=dash line] (64.center) to (59.center);
		\draw [style=dash line] (67.center) to (68.center);
		\draw [style=dash line, bend right=45, looseness=1.25] (68.center) to (69.center);
		\draw [style=dash line] (69.center) to (70.center);
		\draw [style=dash line, bend right=45, looseness=1.25] (70.center) to (71.center);
		\draw [style=dash line] (71.center) to (66.center);
 \end{pgfonlayer}
\end{tikzpicture}
    \caption{A diagrammatic outline of the parallel state shift preparation. We use red lines to mark the substates for increment and blue lines for decrement. The even/odd-decompositon is distinguished with dashed lines for the odd substates. The data is retained in the \emph{amplitudes}: for example, when at the bottom we flip the even and odd parts of the decrement substate, this simply means remapping of the amplitudes. After this preparation step, an inverter on the least significant qubit shifts the states to both directions in parallel.}
    \label{fig:parallel_diagram}
\end{figure*}

The parallelization is done by decomposing the position substate into components involving even and odd basis states. If a basis state $\ket{k}$ is even, the increment can be done by applying an inverter (an $X$ gate in the circuits) to the least significant qubit; and if a basis state is odd, the same inverter defines the decrement. The state decomposition $D$ requires an ancilla qubit $\ket{a}$: for an even state $\ket{k}$, the decomposition maps the amplitude $\alpha_k$ to the state $\ket{k} \otimes \ket{0}_a$ -- while an odd state amplitude is mapped to the state $\ket{k} \otimes \ket{1}_a$ -- where the ancilla qubit is marked with a subscript $a$. After the decomposition, the amplitudes corresponding to the basis states are rearranged in a such a way that applying the inverter on the least significant qubit gives the increment and decrement in parallel. Denoting the rearrangement by $R$ and the inverter by $X$, the shift of an arbitrary linear combination of $N$ basis states is defined by the sequence
\begin{equation}
    \inv{D} X R D \sum_{k=0}^{N-1} \alpha_k \ket{k} \otimes \ket{c} \otimes \ket{a} .
\end{equation}
We have outlined the preparation process up to the inverter in Figure~\ref{fig:parallel_diagram}.

Let us fix the number of positions as $N = 2^{n}$, with $n+1$ being the number of qubits in the working register. Following the usual convention we start both the state and qubit indexing from $0$ so that the last qubit $q_{n}$ in this register can be thought as the coin qubit. The first half $\ket{p_+}$ of the input state vector $\ket{p} = \sum_k \alpha_k \ket{k}$ corresponds to the coin state $\ket{0}_{q_n}$ and should be incremented, while the second half $\ket{p_-}$ corresponding to the coin state $\ket{1}_{q_n}$ should be decremented:
\begin{equation}
    \begin{split}
        \ket{p} & = \ket{p_+} + \ket{p_-} \\
            &= \sum_{k=0}^{2^{n} - 1} \alpha_k \ket{k} \otimes \ket{0}_{q_n} + \sum_{k=2^{n}}^{2^{n+1}-1} \alpha_k \ket{k} \otimes \ket{1}_{q_n} .
    \end{split}
\end{equation}
The distribution of the amplitudes over the increment and decrement substates is dependent on the coin operator, and reflected in the state of the coin qubit.

\begin{figure*}[htbp]
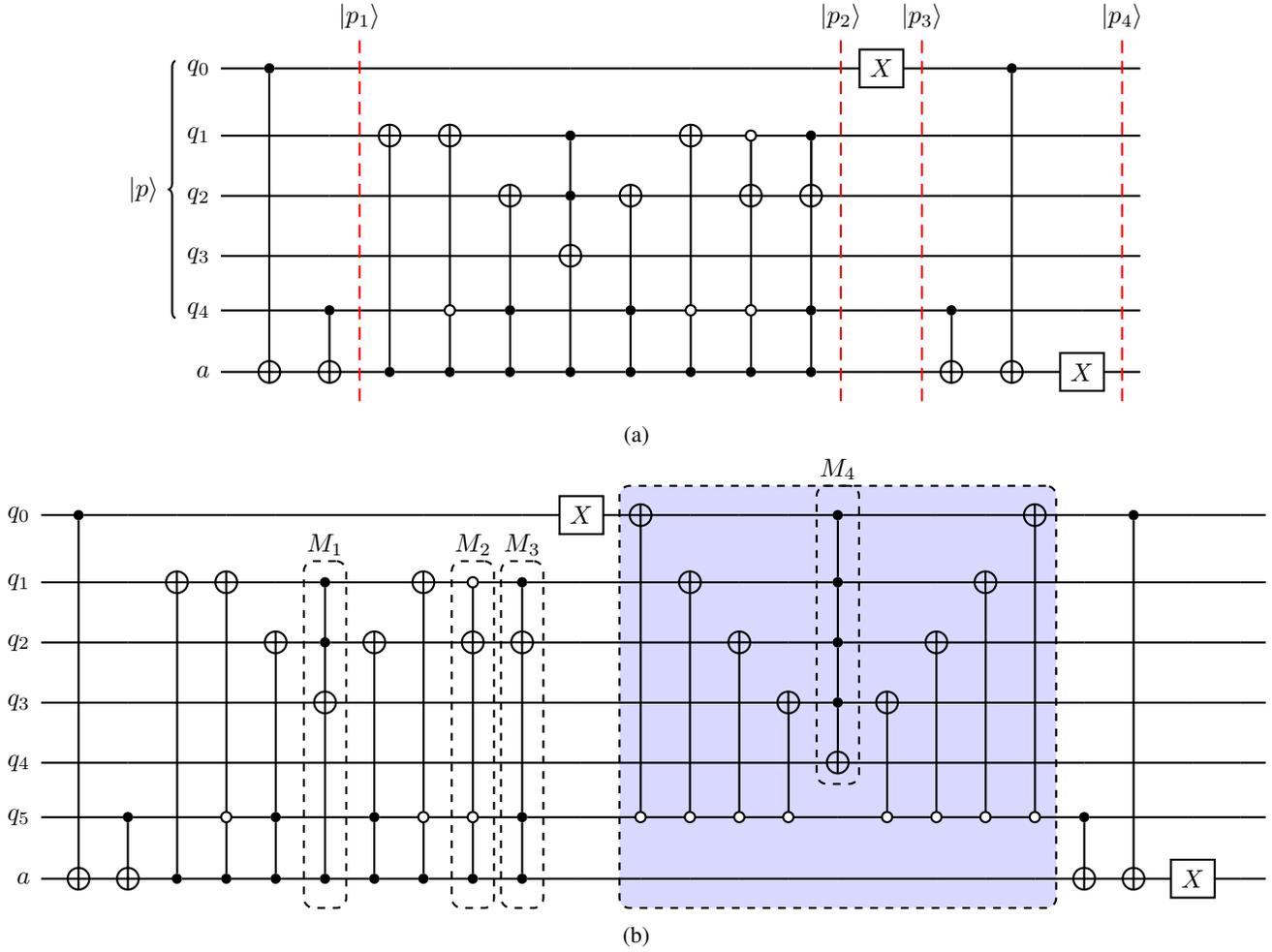

    \centering
    \begin{subfigure}[b]{\linewidth}
        \centering
        \input{circuit-5q}
        \caption{}
        \label{fig:circuit-5q}
    \end{subfigure}
    \hfill
    \begin{subfigure}[b]{\linewidth}
        \centering
        \input{circuit-6q}
        \caption{}
        \label{fig:circuit-6q}
    \end{subfigure}
    \caption{Quantum circuits for the parallel state shift operating on \protect\subref{fig:circuit-5q} five working qubits and \protect\subref{fig:circuit-6q} six working qubits.}
    \label{fig:circuits-5q-6q}
\end{figure*}

Figure~\ref{fig:circuit-5q} illustrates the algorithm using a five-qubit working register $\ket{q}$ and a one-qubit ancilla register $\ket{a}$. In the circuit the first $CX$ gate transforms the input state into a superposition of even and odd parts, and the second $CX$ gate exchanges the even and odd parts of the second half $\ket{p_-}$ of the state vector. This corresponds to the decomposition operator $D$ and we get the result state
\begin{equation}
    \begin{split}
        \ket{p_1} & = D \left(\ket{p} \otimes \ket{a} \right) \\        
        & = \sum_{k(even)=0}^{2^n - 1} \alpha_k \ket{k} \otimes \ket{0}_{q_n} \otimes \ket{0}_a \\
        & + \sum_{k(odd)=2^n}^{2^{n+1}-1} \alpha_k \ket{k} \otimes \ket{1}_{q_n} \otimes \ket{0}_a \\
        & + \sum_{k(odd)=0}^{2^n - 1} \alpha_k \ket{k} \otimes \ket{0}_{q_n} \otimes \ket{1}_a\\
        & + \sum_{k(even)=2^n}^{2^{n+1}-1} \alpha_k \ket{k} \otimes \ket{1}_{q_n} \otimes \ket{1}_a ,
    \end{split}
\end{equation}
where we have, by a slight abuse of notation, differentiated the even and odd basis states.

After this decomposition, applying an $X$ gate on the first qubit $q_0$ results in the right and left shift of the sub-state corresponding to $\ket{0}_a$. However, this will not readily give the shift on the sub-state corresponding to $\ket{1}_a$. Therefore we have to introduce the operator $R$ rearranging the $\ket{1}_a$ sub-state. In the presented circuits, the third $CX$ gate exchanges the amplitudes of neighbouring basis states, while the sequence of the next seven multi-controlled $X$ gates rearranges the corresponding sub-state. This results in a particular structure on which the shift follows from applying the $X$ gate on the first qubit, thus aligning with the sub-state $\ket{0}_a$: for example, this particular structure for the odd states flips the operation of the $X$ gate from the left shift to the right shift. The rearrangement of the state $\ket{p_1}$ then results in the following:
\begin{equation}
    \begin{split}
        \ket{p_2} & = R \left(\ket{p_1} \otimes \ket{a} \right) \\
        & = \sum_{k(even)=0}^{2^n-1} \alpha_{k} \ket{k} \otimes \ket{0}_{q_n} \otimes \ket{0}_a \\
        & + \sum_{k(odd)=2^n}^{2^{n+1}-1} \alpha_k \ket{k} \otimes \ket{1}_{q_n} \otimes \ket{0}_a \\
        & + \sum_{k(odd)=0}^{2^n-1} \alpha_{k-2} \ket{k} \otimes \ket{0}_{q_n} \otimes \ket{1}_a\\
        & + \sum_{k(even)=2^n}^{2^{n+1}-1} \alpha_{k+2} \ket{k} \otimes \ket{1}_{q_n} \otimes \ket{1}_a .
    \end{split}
\end{equation}
This state corresponds to the end state of the diagram in Fig.~\ref{fig:parallel_diagram}. Note that the periodicity defines an equivalence of the $k=0$ and $k=2^{n+1}$ basis states in both odd and even parts. Now the $X$ gate on first qubit $q_0$ increments the first half and decrements the second half of the state vector:
\begin{equation}
    \begin{split}
        \ket{p_3} & = X_{0} \left(\ket{p_2} \otimes \ket{a} \right) \\
        & = \sum_{k(odd)=0}^{2^n-1} \alpha_{k-1} \ket{k} \otimes \ket{0}_{q_n} \otimes \ket{0}_a \\
        & + \sum_{k(even)=2^n}^{2^{n+1}-1} \alpha_{k+1} \ket{k} \otimes \ket{1}_{q_n} \otimes \ket{0}_a \\
        & + \sum_{k(even)=0}^{2^n-1} \alpha_{k-1} \ket{k} \otimes \ket{0}_{q_n} \otimes \ket{1}_a\\
        & + \sum_{k(even)=2^n}^{2^{n+1}-1} \alpha_{k+1} \ket{k} \otimes \ket{1}_{q_n} \otimes \ket{1}_a,
    \end{split}
\end{equation}

The last two $CX$ gates in the circuit combine the sub-states, thus inverting the previous decomposition. The final state $\ket{p_4}$ then has the first half incremented and the second half decremented, as desired. 

In comparison to the canonical state shift, this process implements the two shifts in parallel. On one hand, this can allow to reduce the depth of the circuit. On the other hand, it paves way for efficient shifts in higher dimensions -- in essence, varying the increment and decrement as ``state jumps'' rather than one-step shifts. We will discuss this in more detail in Section~\ref{sec:conclusion}.

\subsection{Scalability}

The presented method has good scalability in terms of the number of qubits. As an example, the circuit for the shift with six working qubits is presented in Figure~\ref{fig:circuit-6q}, where the part of the circuit that differs from the five-qubit circuit of Figure~\ref{fig:circuit-5q} is drawn within a blue box. Here an introduction of one additional qubit into the working register requires a sequence of one multi-controlled $X$ gate sandwiched between four $CX$ gates, while the rest of circuit remains the same. For the sake of simplicity we introduce the new qubit right before the last working qubit -- that is, before the coin qubit. In practical terms this means that the addition of new qubits into the working register changes only the indexing of the coin qubit and the ancilla qubit (that is, their position is incremented by one for each new qubit added). The rest of the register remains at the original position.

\begin{figure*}[htbp]
    \centering
    \input{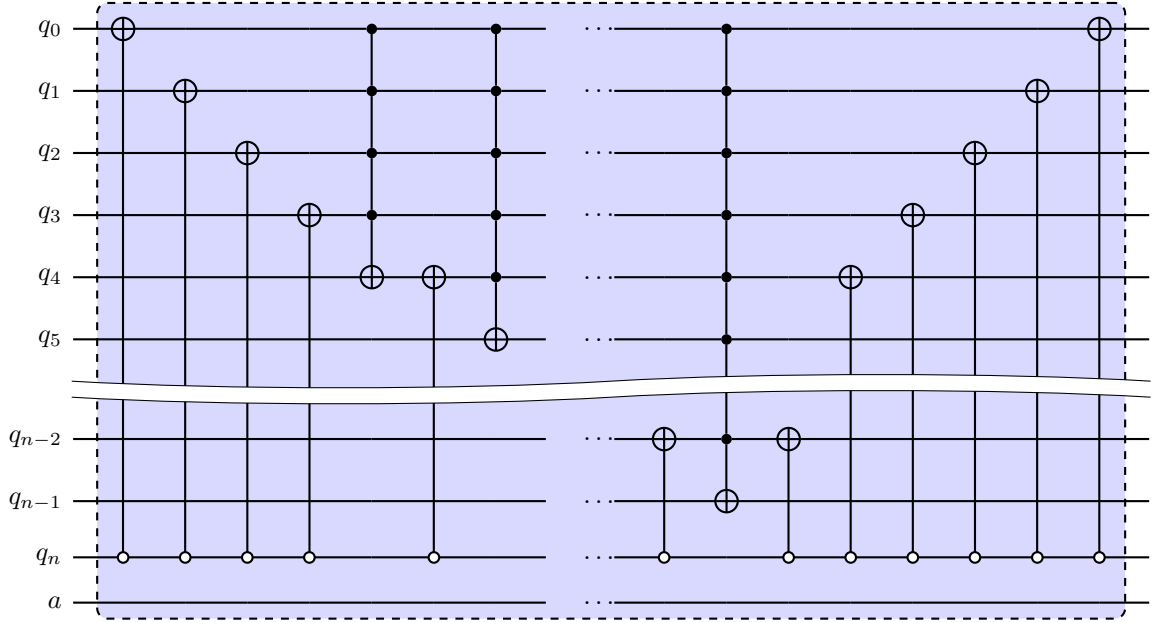}
    \caption{The variable part of the circuit of the parallel shift algorithm.}
    \label{fig:circuit-variable}
\end{figure*}
Going forward to larger working registers, per each added working qubit we introduce one multi-controlled $X$ gate sandwiched between two $CX$ gates, with each new multi-controlled gate requiring one more additional control. This variable part of the circuit is shown in Figure~\ref{fig:circuit-variable}.

\subsection{Decomposition of multi-controlled gates with ancillas}
\label{sec:ancilla_decomposition}

\begin{figure*}[htbp]
    \centering
    \input{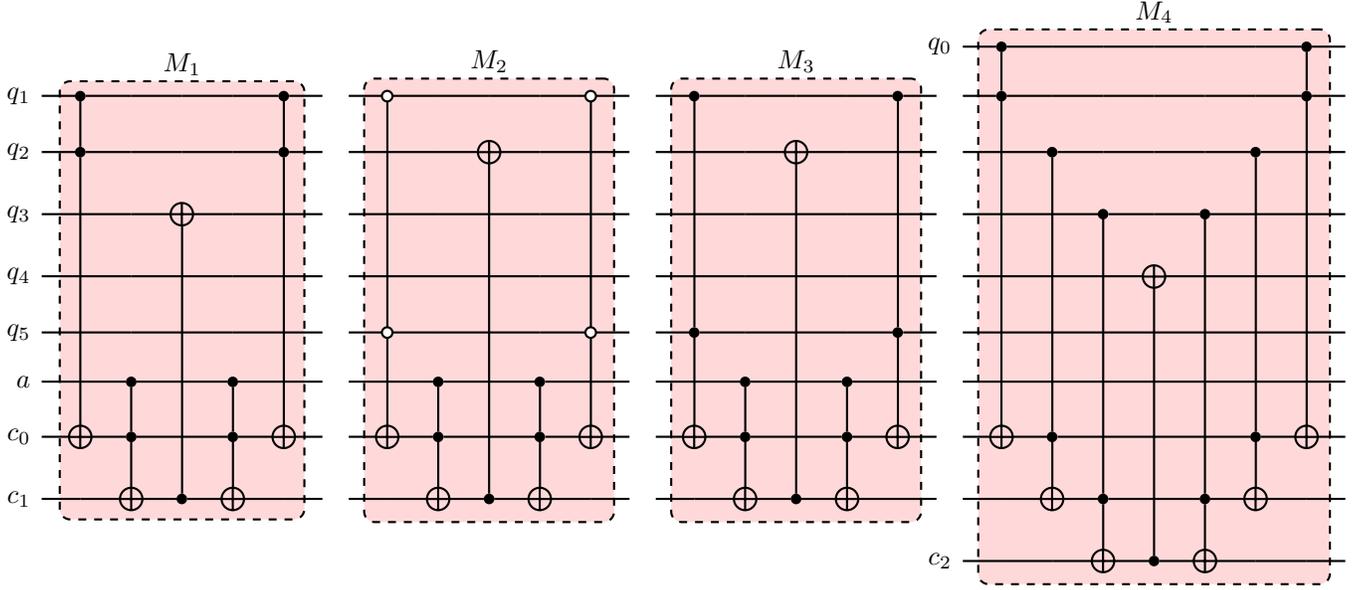}
    \caption{Decompositions of the four multi-controlled gates in Fig.~\protect\ref{fig:circuit-6q} using an extended ancilla register $\ket{c}$.
    }
    \label{fig:circuit-decomposed}
\end{figure*}

As the state shift is based on multi-controlled $X$ gates, the standard decomposition outlined in \cite[pp.~183--184]{Nielsen_Chuang} can be applied. We use this decomposition for multi-controlled $X$ gates having three or more controls, and then the required number of ancilla qubits $m$ depends on the dimension of working register $n$ as $m=n-3$. Each multi-controlled $X$ gate can be decomposed to a sequence of one $CX$ gate sandwiched between two-controlled $X$ gates. Assuming we have a multi-controlled $X$ gate with $n_c$ controls, the number $n_X$ of the two-controlled $X$ gates is given by $n_X= 2(n_c-1)$.
The size of the ancilla register $n_a$ scales linearly with the number of controls as $n_a=n_c-1$. Figure~\ref{fig:circuit-decomposed} shows an example of this decomposition for the multi-controlled gates in the state shift circuit of six working qubits (Figure~\ref{fig:circuit-6q}). 

The decomposition results in a quantum circuit with a linear scaling in terms of the number of gates and the number of qubits required, thus enabling exponential speedup compared to classical shift. As remarked earlier, this is in contrast to the QFT variation which has only quadratic scaling.

\subsection{Towards two-dimensional shifts}

The parallel shift can be expanded into two dimensions, and potentially higher. As an example, we can consider a rectangular lattice with four movement variables assigned to each site. This extends the left- and right-moving state split of the one-dimensional case to a decomposition of four variables, namely left, right, down, and up. The canonical shift provides these four directions again as a sequence, but this can be parallelized by a similar decomposition and rearrangement as in the one-dimensional case.

To apply the parallel shift to all four directions simultaneously, we can introduce a permutation of the grid directions to amount for the up and down shifts through controlled \textsc{SWAP} gates. This adapts the left/right-shift to account for the vertical grid movement. After the permutation step we can follow with the parallel shift as in the one-dimensional case, and to recreate the initial configuration we then revert the permutation as the last step.

In general, the algorithm for two-dimensional shift (four directions) can be visualized as a parallel shift sandwiched between a set of controlled \textsc{SWAP} gates. As with the one-dimensional case, we pay the initial price of a more complex configuration for the amplitude mapping, but gain greatly by the simplicity of the shift itself. This scales well: each added qubit doubling the lattice dimension comes with only a small constant number of additional gates. Remarkably, when comparing to the one-dimensional parallel shift, there is only an initial difference in the gate count stemming from the higher dimension: the scaling coefficient remains exactly same. The need for additional ancilla qubits can be considered reasonable, unless the qubit topology is severely restricted.

Proper exposition of the higher-dimensional parallel shift and applications to practical examples will be presented in future work.

\section{Computational complexity}
\label{sec:complexity}

For real device applications every abstract quantum algorithm needs to decomposed to some set of basis gates. Therefore, it is both appropriate and valuable for practical applications to provide a sense of gate depth on the device level, and analyze the circuit complexity based on such configurations. 

We compare the parallel shift to the canonical and QFT variants using a transpilation with a basis set of gates $CX$, $I$, $RZ$, $SX$ and $X$, which constitute the native IBM gateset. The transpilation is done with the \textsc{Qiskit} SDK using optimization level 1, which maps the circuit to the device configuration and collapses adjacent gates to avoid redundancy.%
~\footnote{See the \href{https://qiskit.org/documentation/apidoc/transpiler.html}{\textsc{Qiskit} transpiler reference} for more information.} 
We assume full qubit connectivity for the sake of clear comparison, although this assumption is not always realistic.

We also do additional manual elimination of redundant gates whenever possible, and apply this for all three state shift methods. This allows us to give a comparative analysis in terms of the number of gates versus the number of qubits; for simplicity, we limit to the range of $5$--$25$ qubits in the working register. Note that in the case of the canonical and the parallel circuit, we use the ancilla decomposition of the multi-controlled gates. For the QFT circuit, there are no multi-controlled gates that could be decomposed with the ancilla register. We stress that exact scaling numbers depend on the use of circuit optimization techniques and the exact target device configuration. Nevertheless, one should see a similar trend regardless of the circuit optimization, as the differences between the three shift algorithms are fundamental.
\begin{figure*}[htbp]
    \centering
    \begin{subfigure}[t]{0.5\textwidth}
        \centering\captionsetup{width=.9\linewidth}
        	\pgfplotstableread{     x             y1           y2           y3           
5	119	105	32
6	171	150	50
7	223	165	72
8	275	180	98
9	327	195	128
10	379	210	162
11	431	225	200
12	483	240	242
13	535	255	288
14	587	270	338
15	639	285	392
16	691	300	450
17	743	315	512
18	795	330	578
19	847	345	648
20	899	360	722
	}{\table}
\begin{tikzpicture}
	\begin{axis}[
	xmin=5, xmax=20,
	ymin=0, ymax=1000,
	xtick distance = 5,
	ytick distance = 200,
	grid = both,
	minor tick num = 1,
	major grid style = {lightgray},
	minor grid style = {lightgray!25},
	width = 0.9\textwidth,
	height = 0.5\textwidth,
	xlabel = {Qubits},
	ylabel = {\#$CX$},legend pos=north west]
	
	\addplot[smooth,thick,blue] table [x = {x}, y = {y1}] {\table};
	\addplot[smooth,thick,red] table [x ={x}, y = {y2}] {\table};
	\addplot[smooth,thick,black] table [x ={x}, y = {y3}] {\table};
	\legend{
		Basic,
		Parallel,
		QFT}
	\end{axis}
\end{tikzpicture}
    	\caption{Comparison of the $CX$ gate scaling between the canonical, the QFT, and the optimized parallel state shift.}
        \label{fig:cnots}
    \end{subfigure}%
    \begin{subfigure}[t]{0.5\textwidth}
        \centering\captionsetup{width=.9\linewidth}
        	\pgfplotstableread{     x             y1           y2           y3       y4           y5           y6      
5	180	154	50	36	28	8
6	260	214	77	52	40	10
7	340	234	110	68	44	12
8	420	254	149	84	48	14
9	500	274	194	100	52	16
10	580	294	245	116	56	18
11	660	314	302	132	60	20
12	740	334	365	148	64	22
13	820	354	428	164	68	24
14	900	374	491	180	72	26
15	980	394	554	196	76	28
16	1060	414	617	212	80	30
17	1140	434	680	228	84	32
18	1220	454	743	244	88	34
19	1300	474	806	260	92	36
20	1380	494	869	276	96	38
	}{\table}
\begin{tikzpicture}
	\begin{axis}[
	xmin=5, xmax=20,
	ymin=0, ymax=1500,
	xtick distance = 5,
	ytick distance = 250, 
	grid = both,
	minor tick num = 1,
	major grid style = {lightgray},
	minor grid style = {lightgray!25},
	width = 0.9\textwidth,
	height = 0.5\textwidth,
	xlabel = {Qubits},
    ylabel = {\#$RZ$ ($-$), \#$SX$ (\Kutline)},
    legend pos=north west]
	
	\addplot[smooth,thick,blue] table [x = {x}, y = {y1}] {\table};
	\addplot[smooth,thick,red] table [x ={x}, y = {y2}] {\table};
	\addplot[smooth,thick,black] table [x ={x}, y = {y3}] {\table};
    \addplot[dashed,thick,blue] table [x = {x}, y = {y4}] {\table};
	\addplot[dashed,thick,red] table [x ={x}, y = {y5}] {\table};
	\addplot[dashed,thick,black] table [x ={x}, y = {y6}] {\table};
	\legend{
		Basic,
		Parallel,
		QFT}
	\end{axis}
\end{tikzpicture}
    	\caption{Comparison of the $RZ$ and $SX$ gate scaling between the canonical, the QFT, and the optimized parallel state shift.}
        \label{fig:singlegates}
    \end{subfigure}
    \caption{A comparison of basis gate scaling between the three variations of the state shift. The number of qubits is the size of the working register without ancilla.}
\end{figure*}
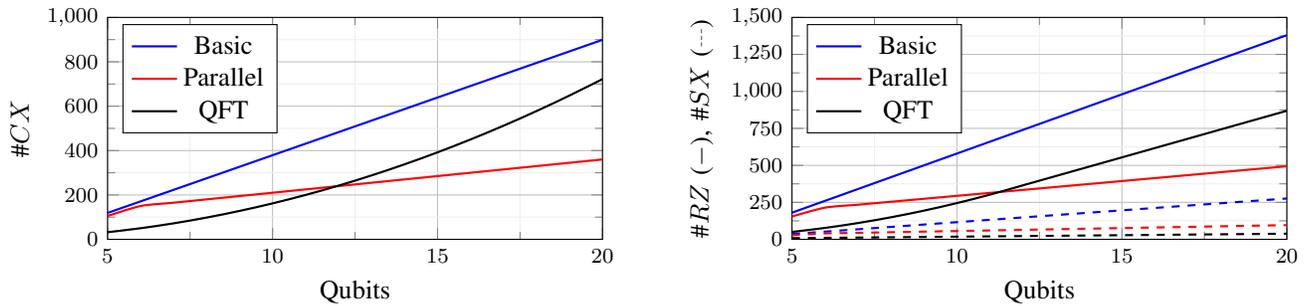 

Figure~\ref{fig:cnots} shows the scaling of the $CX$ gates with the number of the working qubits for all three methods, and in Figure~\ref{fig:singlegates} we have the same comparison for the $RZ$ and $SX$ gates. As can be seen, the scaling behaviour is linear for all the quantum gates in the cases of the parallel and the canonical shift, while the QFT circuit has a quadratic dependence.  However, there is a significant improvement from the canonical to the optimized parallel algorithm in terms of the $CX$-gate scaling. In the case of the canonical shift, the number of $CX$ gates scales with the number of working qubits as $52n - 141$, while after parallelization the scaling can be given as $15n + 59$. The improvement factor defined as the slope ratio between the scaling curves of these two algorithms is $3.5$.

On the other hand, the $CX$ gates for the QFT circuit scale quadratically with the number of working qubits as $2n^2 - 4n + 2$. Due to the lower number of initial $CX$ gates, the QFT method provides smaller circuits up to $12$ working qubits, after which the quadratic scaling makes it inferior in comparison to the parallel shift. In practice, the exact tipping point depends on the qubit connectivity and the transpilation process. For example, with the IBM Falcon architecture this improvement becomes tangible at $14$ working qubits. 
\begin{table}[htbp]
    \centering
    \caption{The $CX$ gate count comparison of the QFT and the parallel shift, using different \textsc{tket} optimization levels and varying the size of the working register.}
    \label{tab:cnot_count}
    \begin{tabular}{@{}llrr@{}}
        \toprule
        \#Qubits & Optimization & \multicolumn{2}{c}{\#CX} \\ \cmidrule{1-1}  \cmidrule{2-2} \cmidrule{3-4}
        & & QFT & Parallel \\ \cmidrule(lr){3-3} \cmidrule(l){4-4}
        \multirow{2}{*}{10} & 1 & 162 & 209 \\
        & 2 & 159 & 206 \\ \midrule
        \multirow{2}{*}{15} & 1 & 392 & 284 \\
        & 2 & 389 & 281 \\ \midrule
        \multirow{2}{*}{20} & 1 & 722 & 359 \\
        & 2 & 719 & 356 \\ \midrule
        \multirow{2}{*}{25} & 1 & 1152 & 434 \\
        & 2 & 1149 & 431 \\
        \bottomrule
    \end{tabular}
\end{table}

In the case of $RZ$ gates there is a linear scaling $80n - 220$ and $20n + 94$ for the canonical and the parallel state shift, respectively. The QFT again has a quadratic dependence of $n^2 + 34n - 177$. For $SX$ gates, all three methods have linear scaling.

For concrete numbers of $CX$ gates we provide the Table~\ref{tab:cnot_count}. Here we have used the \textsc{tket} compiler with two predefined optimization sequences (optimization levels $1$ and $2$) via the \textsc{pytket} package. The level $1$ is described as preserving the qubit connections and the target gateset but removing redundancies. The level $2$ introduces further Clifford simplifications, commutes single-qubit gates to the front of the circuit and attempts to squash local subcircuits.%
~\footnote{More details can be found in the \href{https://cqcl.github.io/pytket/manual/index.html}{\textsc{pytket} manual}.}
The benefit of using \textsc{tket} over \textsc{Qiskit} transpiler is in having a deterministic compilation sequence, as higher optimization levels in \textsc{Qiskit} have stochastic behaviour which would be an unnecessary complication for this comparison. By using the default compilation passes we avoid tailoring the sequence for these specific circuits.

\subsection{Analysis of the \texorpdfstring{$CX$}{CX} %
    gate complexity}

The examples above illustrate the efficiency of the parallel shift in wider circuits. As the parallelization algorithm relies heavily on multi-controlled gates, it is also appropriate to provide a general account on the resulting complexity. In the following, we denote an $k$-controlled $X$-gate by $C^{k}X$, and skip the effect of the single-qubit gates. Lower $CX$ number is crucial especially for the early applications, as the number of two-qubit gates is often the limiting factor for obtaining meaningful results from a real device.

From Figure~\ref{fig:circuits-5q-6q} we can gather the \emph{constant} and the \emph{variable} parts of the shift circuit. For the constant part, we have a fixed number of multi-controlled gates with a total of $5$ $CX$ gates, $4$ $C^2X$ gates, and $3$ $C^3X$ gates. These figures remain constant regardless of the size of the working register, and enable us to do the decompositon and the parallel shift of the decomposed substates. The variable part is needed for circuits with a working register larger than five qubits; this part accounts for the rearrangement of the shifted amplitudes back to the original configuration. If the number of qubits in the working register is $n$, this part includes $2(n-2)$ $CX$ gates and a sequence $\left\{C^{k}X\right\}_{k=4}^{n-2}$ of multi-controlled gates.

The total count in $CX$ gates now depends on how the multi-controlled gates are to be decomposed. The lower limit for a decomposition of $C^{k}X$ gate is $2(k+1)$ $CX$ gates without introducing an additional ancilla register~\cite{shende2009}. In general we should attach a scaling factor $d_k$ to each $C^kX$ gate, giving the total $CX$ gate count as
\begin{equation}
    n_{CX}(n) = 1 + 2n + 4d_2 + 3d_3 + \sum_{k=4}^{n-2} d_k .
\end{equation}
The lower limit is then $49 + 2n + \sum_{k=4}^{n-2}2(k+1)$ $CX$ gates.

In Section~\ref{sec:ancilla_decomposition} we outlined the decomposition of $C^kX$ gates into sequences of $CX$ and Toffoli ($C^2X$) gates with the help of an ancilla register. For each $C^kX$ we get a sequence of $2(k-1)$ Toffoli gates and a single $CX$ gate: in this configuration the problem then reduces to the decomposition of Toffoli gates, for which the optimal number of $CX$ gates is known to be six~\cite{shende2009}. From this we can gather the total amount of $CX$ gates with respect to the size $n > 5$ of the working register as
\begin{equation}
\label{eq:cnots_dumb}
    n_{CX}(n) = 96 + 3n + \sum_{k=4}^{n-2} 12(k-1) ,
\end{equation}
where the constant part of the shift circuit amounts to the total of $104$ $CX$ gates.
\begin{figure*}[htbp]
    \centering
    \input{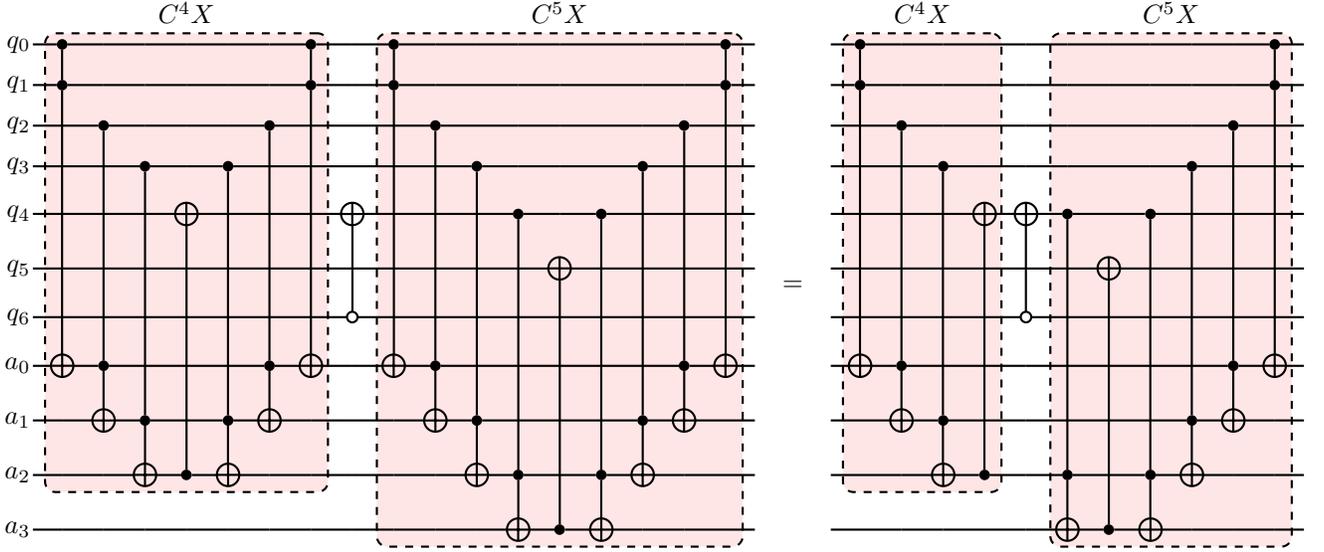}
    \caption{Removing redundancies in the variable part of the parallel shift circuit: an example of sequenced $C^4X$ and $C^5X$ gates in the case of $n = 7$ qubits in the working register. The generalization to longer sequences is straightforward.}
    \label{fig:circuit-toffoli_cancellation}
\end{figure*}

This is not the whole story, however. These figures follow from applying the decomposition for each of the multi-controlled gates separately, and then summing up the results. As the gates are systematically placed into the configuration outlined in Figure~\ref{fig:circuit-variable}, we can exploit the emerging redundancies due to the Toffoli gate being its own inverse. In the variable part we have pairs of $C^kX$ and $C^{k+1}X$ gates with a single $CX$ gate in-between, with the controls placed in these gates in such a way that the right-hand side of the Toffoli cascade of the decomposed $C^k$ gate cancels out entirely, leaving one Toffoli gate on the left-hand side of the Toffoli cascade of the gate $C^{k+1}X$; see Figure~\ref{fig:circuit-toffoli_cancellation} for an example with $C^4X$ and $C^5X$ gates in the sequence. We can remove $2(k-1)$ Toffoli gates from the pair, leaving a total of $k + 1$ Toffoli gates and two $CX$ gates (ignoring the one $CX$ gate in-between). As this compounds, the number of $CX$ gates will be greatly reduced. If we have six qubits in the working register, the total count is $n_{CX}(6) = 149$, and in this configuration there are no reductions. However, each additional working qubit adds only three $CX$ gates and two Toffoli gates -- meaning $15$ additional $CX$ gates -- amounting to the total
\begin{equation}
\label{eq:cnots_cancelled}
    n_{CX}(n) = 15(n-6) + 149,
\end{equation}
for the size $n>5$ of the working register. Thus we have recovered the linear scaling we noted earlier in this section. In terms of the grid size of $N = 2^m$ positions this leads to a circuit with $15m + 74$ $CX$ gates.

\section{Applications}
\label{sec:applications}

The state shift is a building block for several quantum algorithms, from simulating a quantum walk as a part of the variety of quantum search and simulation models~\cite{Shakeel, Childs2}, to encoding matrices~\cite{Daan} and implementing the propagation step in the lattice Boltzmann models~\cite{Budinski1, Blaga, Budinski2}. We will outline these three applications next and discuss how all of them can benefit from the optimized parallel shift.       

\subsection{Discrete quantum walk}

A discrete-time quantum walk is a quantum analogue of the classical random walk where at each time step walker is moving one step in each direction on an array of discrete sites conditioned by a scattering process. For simplicity, let us consider a one-dimensional cyclical array of $N$ discrete sites with a one-qubit control state. The state of the walker system can then be described by a product state
\begin{equation}
    \ket{\Psi}= \ket{p} \otimes \ket{c}
\end{equation}
where $\ket{c} \in \Hs_c$ is the one-qubit coin state, and the $\ket{p} \in \Hs_p$ is the position state of dimension $N$. The Hilbert space of states is then simply $\Hs=\Hs_c \otimes \Hs_p \coloneqq \Cs^2 \otimes \Cs^N$.

Quantum walk begins with a coin operator $C$, defined as an unitary acting on the coin state -- a Hadamard operator $H$ is often used for one-dimensional quantum walks to create a symmetric superposition. The position state $\ket{p}$ is then shifted with an operator $S$, conditioned by the coin state $\ket{c}$. The shift operator is formally defined as 
\begin{equation}
    S = \ket{0}\bra{0} \otimes \sum_{k=1}^{N} \ket{k-1} \bra{k} + \ket{1}\bra{1} \otimes \sum_{k=1}^{N} \ket{k+1} \bra{k} .
\end{equation}
The coin state then defines incremental (coin component in basis $\ket{1}$) and decremental (coin component in basis $\ket{0}$) movements relative to its linear combination of the basis states. The action of the shift operator $S$ on the state $\ket{p}$ can be parallelized using the sequence $\inv{D} X R D$ described in Section~\ref{sec:parallel}.      

\subsection{Quantum lattice Boltzmann method}

The lattice Boltzmann method (LBM) \cite{Rivet,Rothman,Chen_1998} is an alternative approach to the classical multiphysics simulation for fluid flow. To leverage quantum computing for multiphysics, quantum lattice Boltzmann method (QLBM) has been recently developed for the advection-diffusion and the incompressible Navier-Stokes equations~\cite{Budinski1,Budinski2}. Other approaches can be found in \cite{Itani_Succi_2022, schalkers2022efficient,schalkers2023importance,li2023potential,yamazaki2023quantum}. In contrast to macroscale methods, the LBM replaces fluid density with probability distributions of fictive particles. These are then subjected to propagation and collision processes over a discrete lattice covering the fluid domain. 

\begin{figure}[!htb]
    \centering
    \begin{tikzpicture}

    \fill (0,0) circle [radius=1.75pt] node[label=below:{$i=0$}] (0) {};
    \fill (2.5,0) circle [radius=1.75pt] node[label=below:{$i=1$}] (1) {};
    \fill (5,0) circle [radius=1.75pt] node[label=below:{$i=2$}] (2) {};
    \fill (7.5,0) circle [radius=1.75pt] node[label=below:{$i=N-1$}] (N) {};

    \node (dots) at (6.75,0) {\dots};
    \draw (0.center) -- (dots);
    \draw (dots) -- (N.center);
    
    \draw[->, >=latex, shorten >= 1.25cm, very thick] (1.center) node[fill=white,left = 0.3,inner sep=1pt]{$2$} -- (0);
    \draw[->, >=latex, shorten >= 1.25cm, very thick] (1.center) node[fill=white,right = 0.3,inner sep=1pt]{$1$} -- (2);
    \draw[->, >=latex, shorten >= 1.25cm, very thick] (2.center) node[fill=white,left = 0.3,inner sep=1pt]{$2$} -- (1);
    \draw[->, >=latex, shorten >= 1.25cm, very thick] (2.center) node[fill=white,right = 0.3,inner sep=1pt]{$1$} -- (N);

\end{tikzpicture}
    \caption{D1Q2 lattice configuration for the LBM.}
    \label{fig:boltzmann}
\end{figure}
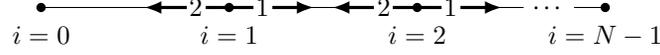

A single relaxation time lattice Boltzmann equation consists of three steps. First there is \emph{collision}:
    \begin{equation}
        \hat{f}_{\alpha}\left(\mathbf{x},t\right)=\left(1-\omega\right) f_{\alpha}\left(\mathbf{x},t\right)+\omega f_{\alpha}^{eq},
    \end{equation}
followed by \emph{streaming}:
    \begin{equation}
        f_{\alpha}\left(\mathbf{x}+\mathbf{e}_{\alpha}\Delta t,t+\Delta t \right) = \hat{f}_{\alpha},
    \end{equation}
and finally the calculation of \emph{macroscopic quantities}:
    \begin{equation}
        \phi_{\alpha}\left(\mathbf{x},t\right) = \sum_{\alpha} f_{\alpha}.
    \end{equation}
Here $f_{\alpha}$ is the particle distribution function along the $\alpha$ link, $f_{\alpha}^{eq}$ is the local equilibrium distribution function, $\mathbf{e}_{\alpha}$ is the particle velocity vector, $\mathbf{x}$ is the location vector defined in the Cartesian coordinates, $t$ is time, and $\omega$ is the single relaxation time. Figure~\ref{fig:boltzmann} illustrates an example of a one-dimensional lattice with two velocity vectors $\mathbf{e}_1 = -\mathbf{e}_2$ for distribution functions $f_1$ and $f_2$, respectively, streaming along the links (line connecting two lattice sites) in opposite directions.

QLBM encodes the distribution functions $f_{\alpha}$ into the state amplitudes, and the spatial position of the lattice sites is encoded into the computational basis states $\ket{k}$. In the case of a simple 1D QLBM model as in Figure~\ref{fig:boltzmann}, this means that the vectors containing two distribution functions $f_1$ and $f_2$ (each having a dimension $N$) are encoded into the quantum system as a superposition of two sub-states. As a consequence of such encoding, the \emph{streaming} step can be efficiently conducted by the state shift. The propagation of the distribution functions along the lattice at each time step follows from applying one-step movement of the distribution function $f_1$ to the right and $f_2$ to the left. 

In contrast to the canonical state shift used in~\cite{Budinski1} and \cite{Budinski2}, the parallel shift significantly improves the propagation step, enhancing the efficiency of multiphysics simulations on quantum devices. For the sake of example, we could reach simulations of the order of $10^9$ lattice points using only $60$ qubits in total, resulting in a $CX$ gate count of roughly $600$ per time step for the propagation step. A simulation of $10^{12}$ lattice points could be carried out with approximately $80$ total qubits and $750$ $CX$ gates. This unlocks the possibility of solving fluid dynamics problems of meaningful size using intermediate-scale quantum computers.

\subsection{Matrix encoding}

\begin{figure*}[!ht]
    \centering
    \input{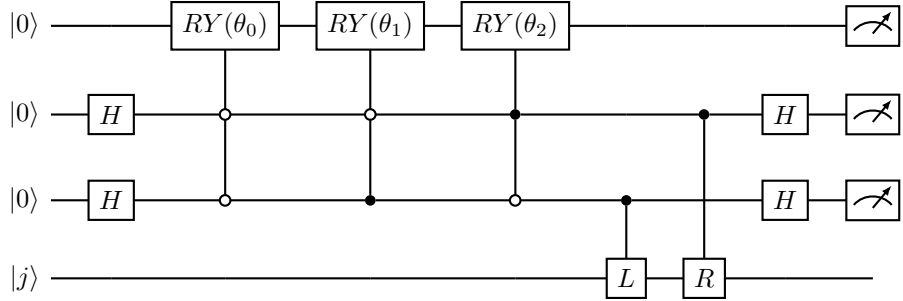}
    \caption{A circuit for the block encoding of a $8\times8$ circulant matrix, adapted from \cite{Daan}.}
    \label{fig:block_encoding}
\end{figure*}

Solving general linear algebra problems with a quantum computer requires a method for encoding non-unitary matrices into the quantum system. The standard way to do this is the block-encoding method, in which a properly scaled non-unitary matrix is embedded into a larger unitary which can then be implemented as a quantum circuit. As reported in \cite{Daan}, efficient implementations of the method are rather restricted: generally, this works only for some well-structured sparse matrices. 

The state shift is an essential part of each block-encoding subroutine. In Figure~\ref{fig:block_encoding} we have an example of a quantum circuit for encoding a banded circulant adjacency matrix of dimension $8\times8$ as defined in \cite{Daan}. In this circuit, the operators $L$ and $R$ correspond to the left and right shift, respectively. The same operators can also be used for encoding the adjacency matrix for an undirected and balanced binary tree, as well as symmetric stochastic matrices. 

\section{Discussion and further work}
\label{sec:conclusion}

In this work we have presented a new algorithm for the one-dimensional, periodic quantum basis state shift. Exploiting the independence of the shift directions, we introduced a parallelization of the canonical shift based on multi-controlled gates. Our method divides the input state into even and odd basis components, which allows to execute the increment and decrement shifts in parallel. While this requires a more complex state setup, overall we gain a lower number of quantum gates in contrast to the other known shift methods.

A complexity analysis highlights the benefits of the parallelization in comparison to the canonical and the QFT-based shifts, when looking at the number of quantum gates against the size of the working qubit register. To give a clear picture of this scaling, we have presented concrete examples of transpiled circuit depth as well as detailed analysis of the gate count. Most notably, we have shown that in terms of two-qubit $CX$ gates, the quadratic scaling of the QFT shift is inferior to the linear scaling of the parallel shift when the size of the working register is at least $12$ qubits. While the canonical shift too scales linearly, it has significant overhead compared to the parallel shift. In all of these comparisons we have assumed that the multi-controlled gates are decomposed with an ancilla register, whenever applicable.

One-dimensional shifts are not the end of the story. We can extend the state shift to other applications by interpreting states as nodes in a higher-dimensional graph or lattice. For example, on a two-dimensional regular lattice we would have not only the directions left and right, but also up and down. This is relevant for example when considering a discrete velocity field of a simple two-dimensional D2Q5 lattice Boltzmann model, which is based on four discrete non-zero velocities for each lattice site. %
%
The parallelization method can be used in these higher-dimensional shifts. For example, instead of having a sequence of left, right, up, and down shifts, the parallel shift allows to consider these directions in effect simultaneously. More details on this extension will be provided in an upcoming article.

\section*{Data availability}
No new data were created or analysed in this study.

\bibliographystyle{IEEEtran}
\bibliography{Main}

\begin{thebibliography}{10}
\providecommand{\url}[1]{#1}
\csname url@samestyle\endcsname
\providecommand{\newblock}{\relax}
\providecommand{\bibinfo}[2]{#2}
\providecommand{\BIBentrySTDinterwordspacing}{\spaceskip=0pt\relax}
\providecommand{\BIBentryALTinterwordstretchfactor}{4}
\providecommand{\BIBentryALTinterwordspacing}{\spaceskip=\fontdimen2\font plus
\BIBentryALTinterwordstretchfactor\fontdimen3\font minus
  \fontdimen4\font\relax}
\providecommand{\BIBforeignlanguage}[2]{{%
\expandafter\ifx\csname l@#1\endcsname\relax
\typeout{** WARNING: IEEEtran.bst: No hyphenation pattern has been}%
\typeout{** loaded for the language `#1'. Using the pattern for}%
\typeout{** the default language instead.}%
\else
\language=\csname l@#1\endcsname
\fi
#2}}
\providecommand{\BIBdecl}{\relax}
\BIBdecl

\bibitem{Venegas_Andraca_2012}
\BIBentryALTinterwordspacing
S.~E. Venegas-Andraca, ``Quantum walks: a comprehensive review,'' \emph{Quantum
  Information Processing}, vol.~11, no.~5, pp. 1015--1106, 2012. [Online].
  Available: \url{https://doi.org/10.1007/s11128-012-0432-5}
\BIBentrySTDinterwordspacing

\bibitem{Douglas_Wang}
\BIBentryALTinterwordspacing
B.~Douglas and J.~Wang, ``Efficient quantum circuit implementation of quantum
  walks,'' \emph{Physical Review A}, vol.~79, no.~5, p. 052335, 2009. [Online].
  Available: \url{https://doi.org/10.1103/physreva.79.052335}
\BIBentrySTDinterwordspacing

\bibitem{Shakeel}
\BIBentryALTinterwordspacing
A.~Shakeel, ``Efficient and scalable quantum walk algorithms via the quantum
  fourier transform,'' \emph{Quantum Information Processing}, vol.~19, no.~9,
  p. 323, 2020. [Online]. Available:
  \url{https://doi.org/10.1007/s11128-020-02834-y}
\BIBentrySTDinterwordspacing

\bibitem{Nielsen_Chuang}
\BIBentryALTinterwordspacing
M.~A. Nielsen and I.~L. Chuang, \emph{{Quantum Computation and Quantum
  Information: 10th Anniversary Edition}}.\hskip 1em plus 0.5em minus
  0.4em\relax Cambridge University Press, Cambridge, 2010. [Online]. Available:
  \url{https://doi.org/10.1017/cbo9780511976667}
\BIBentrySTDinterwordspacing

\bibitem{shende2009}
\BIBentryALTinterwordspacing
V.~V. Shende and I.~L. Markov, ``On the cnot-cost of toffoli gates,''
  \emph{Quantum Information \& Computation}, vol.~9, no.~5, pp. 461--486, 2009.
  [Online]. Available: \url{https://dl.acm.org/doi/10.5555/2011791.2011799}
\BIBentrySTDinterwordspacing

\bibitem{Childs2}
\BIBentryALTinterwordspacing
A.~M. Childs, ``Universal computation by quantum walk,'' \emph{Physical Review
  Letters}, vol. 102, no.~18, p. 180501, 2009. [Online]. Available:
  \url{https://doi.org/10.1103/physrevlett.102.180501}
\BIBentrySTDinterwordspacing

\bibitem{Daan}
\BIBentryALTinterwordspacing
D.~Camps, L.~Lin, V.~B. Roel, and C.~Yang, ``Explicit quantum circuits for
  block encodings of certain sparse matrices,'' 2022, arXiv preprint. [Online].
  Available: \url{https://doi.org/10.48550/ARXIV.2203.10236}
\BIBentrySTDinterwordspacing

\bibitem{Budinski1}
\BIBentryALTinterwordspacing
{\relax Lj}.~Budinski, ``Quantum algorithm for the advection–diffusion
  equation simulated with the lattice boltzmann method,'' \emph{Quantum
  Information Processing}, vol.~20, no.~2, p.~57, 2021. [Online]. Available:
  \url{https://doi.org/10.1007/s11128-021-02996-3}
\BIBentrySTDinterwordspacing

\bibitem{Blaga}
\BIBentryALTinterwordspacing
B.~N. Todorova and R.~Steijl, ``Quantum algorithm for the collisionless
  boltzmann equation,'' \emph{Journal of Computational Physics}, vol. 409, p.
  109347, 2020. [Online]. Available:
  \url{https://doi.org/10.1016/j.jcp.2020.109347}
\BIBentrySTDinterwordspacing

\bibitem{Budinski2}
\BIBentryALTinterwordspacing
{\relax Lj}.~Budinski, ``Quantum algorithm for the {N}avier–{S}tokes
  equations by using the streamfunction-vorticity formulation and the lattice
  {B}oltzmann method,'' \emph{International Journal of Quantum Information},
  vol.~20, no.~2, p. 2150039, 2022. [Online]. Available:
  \url{https://doi.org/10.1142/s0219749921500398}
\BIBentrySTDinterwordspacing

\bibitem{Rivet}
\BIBentryALTinterwordspacing
J.~P. Rivet and J.~P. Boon, \emph{Lattice Gas Hydrodynamics}.\hskip 1em plus
  0.5em minus 0.4em\relax Cambridge University Press, London, 2001. [Online].
  Available: \url{https://doi.org/10.1017/cbo9780511524707}
\BIBentrySTDinterwordspacing

\bibitem{Rothman}
\BIBentryALTinterwordspacing
D.~H. Rothman and S.~Zaleski, \emph{Lattice-Gas Cellular Automata – Simple
  Models of Complex Hydrodynamics}.\hskip 1em plus 0.5em minus 0.4em\relax
  Cambridge University Press, London, 1996. [Online]. Available:
  \url{https://doi.org/10.1017/cbo9780511524714}
\BIBentrySTDinterwordspacing

\bibitem{Chen_1998}
\BIBentryALTinterwordspacing
S.~Chen and G.~D. Doolen, ``Lattice boltzmann method for fluid flows,''
  \emph{Annual Review of Fluid Mechanics}, vol.~30, no.~1, pp. 329--364, 1998.
  [Online]. Available: \url{https://doi.org/10.1146/annurev.fluid.30.1.329}
\BIBentrySTDinterwordspacing

\bibitem{Itani_Succi_2022}
\BIBentryALTinterwordspacing
W.~Itani and S.~Succi, ``Analysis of carleman linearization of lattice
  boltzmann,'' \emph{Fluids}, vol.~7, no.~1, p.~24, 2022. [Online]. Available:
  \url{https://doi.org/10.3390/fluids7010024}
\BIBentrySTDinterwordspacing

\bibitem{schalkers2022efficient}
\BIBentryALTinterwordspacing
M.~A. Schalkers and M.~{\relax M}öller, ``Efficient and fail-safe
  collisionless quantum boltzmann method,'' 2022, arXiv preprint. [Online].
  Available: \url{https://doi.org/10.48550/arXiv.2211.14269}
\BIBentrySTDinterwordspacing

\bibitem{schalkers2023importance}
\BIBentryALTinterwordspacing
{Schalkers, M. A. and {\relax M}öller, M.}, ``On the importance of data
  encoding in quantum boltzmann methods,'' 2023, arXiv preprint. [Online].
  Available: \url{https://doi.org/10.48550/arXiv.2302.05305}
\BIBentrySTDinterwordspacing

\bibitem{li2023potential}
\BIBentryALTinterwordspacing
X.~Li, X.~Yin, N.~Wiebe, J.~Chun, G.~K. Schenter, M.~S. Cheung, and
  J.~Mülmenstädt, ``Potential quantum advantage for simulation of fluid
  dynamics,'' 2023, arXiv preprint. [Online]. Available:
  \url{https://doi.org/10.48550/arXiv.2303.16550}
\BIBentrySTDinterwordspacing

\bibitem{yamazaki2023quantum}
\BIBentryALTinterwordspacing
S.~Yamazaki, F.~Uchida, K.~Fujisawa, and N.~Yoshida, ``Quantum algorithm for
  collisionless boltzmann simulation of self-gravitating systems,'' 2023, arXiv
  preprint. [Online]. Available:
  \url{https://doi.org/10.48550/arXiv.2303.16490}
\BIBentrySTDinterwordspacing

\end{thebibliography}

\end{document}